\documentclass[preprint,12pt]{elsarticle}

\usepackage{amssymb}

\journal{Nuclear Physics B}

\begin{document}

\begin{frontmatter}



\title{Charmonium production in ultra-peripheral heavy ion collisions with two-photon processes}







\author[a]{Gong-Ming Yu\corref{mycorrespondingauthor}}
\cortext[mycorrespondingauthor]{Corresponding author}
\ead{ygmanan@163.com}


\author[a,b]{Yue-Chao Yu}


\author[b]{Yun-De Li}

\author[a]{Jian-Song Wang\corref{mycorrespondingauthor}}
\ead{jswang@impcas.ac.cn}

\address[a]{Institute of Modern Physics, Chinese Academy of Sciences, Lanzhou 730000, China}

\address[b]{Department of Physics, Yunnan University, Kunming 650091, China}


\begin{abstract}
We calculate the production of large-$p_{T}$ charmonium and narrow resonance state (exotic charmonium) in proton-proton,
proton-nucleus, and nucleus-nucleus collisions with the semi-coherent two-photon interactions at
Relativistic Heavy Ion Collider (RHIC), Large Hadron Collider (LHC), and Future Circular Collider (FCC) energies.
Using the large quasi-real photon fluxes, we present the $\gamma\gamma\rightarrow H$ differential cross section for
charmonium and narrow resonance state production at large transverse momentum in ultra-peripheral heavy ion collisions. The numerical results
demonstrate that the experimental study of ultra-peripheral collisions is feasible at RHIC, LHC, and FCC energies.


\end{abstract}






\end{frontmatter}



\section{Introduction}

The two-photon interactions\cite{1,2,3,4,5,6} at unprecedentedly high energies can be studied in ultra-peripheral heavy ion collisions. Recently, the measurement for ultra-peripheral collisions was performed by PHENIX Collaboration\cite{6x0} and STAR Collaboration\cite{6x01,6x02} at Relativistic Heavy Ion Collider (RHIC) energies, as well as ALICE Collaboration\cite{6x1,6x2,6x20,6x21,6x22,6x23} and CMS Collaboration\cite{6x3,6x4,6x5} at Large Hadron Collider (LHC) energies. Moreover, the equivalent photon flux present in high energy nuclear collisions
is very high, and has been found many useful applications in photon production\cite{7,8,9,10,11}, dilepton production\cite{12,13,14,15,15x,16,17}, jet production\cite{18,19}, double vector meson production\cite{20,21,22,22x}, (exotic) charmonium production\cite{23,24,25,26,27,28}, decuplet-baryon pair production\cite{29}, doubly heavy baryon production\cite{29x}, graviton production\cite{30}, $W/Z$ boson production\cite{31,32}, and Higgs boson production\cite{33,34,35} in the ultra-peripheral heavy ion collisions. The most existing calculations concentrated on the total cross section and also of the rapidity or the two-photon energy distributions, an interesting theoretical quantity, which may not be measured easily in ultra-peripheral collisions at Relativistic Heavy Ion Collider (RHIC), Large Hadron Collider (LHC), and Future Circular Collider (FCC), since the experiments running at RHIC, LHC, and those planned at FCC demand the distributions of transverse momenta and rapidity.

In the present work, we investigate the semi-coherent two-photon processes for large-$p_{T}$ charmonium and narrow resonance state (exotic charmonium) production in proton-proton, proton-nucleus, and nucleus-nucleus collisions at RHIC, LHC, and FCC energies. If the transverse momenta of both photons are the same large (non-coherent) or small (coherent), the total transverse momentum would have a very small value, then the charmonium and narrow resonance state (exotic charmonium) could not obtain large transverse momentum in the $\gamma\gamma\rightarrow H$ interaction, where $H$ is the (exotic) charmonium. Indeed, the single track condition \cite{15x,16} leads to the weak contribution of non-coherent and coherent photon-photon processes for large-$p_{T}$ charmonium and narrow resonance state (exotic charmonium) production compared with the semi-coherent processes. Hence we work in the semi-coherent two-photon approach, in which one photon is relatively hard and is incoherently emitted by participating projectile, while another can be soft enough to be in a coherent domain. In the equivalent photon approximation that the effect of the electromagnetic field for an ultrarelativistic proton or nucleus is replaced by a flux of photons, we present the differential cross section as a function of the transverse momenta for the large-$p_{T}$ charmonium, as well as the narrow resonance state (exotic charmonium)\cite{36,37,38,39} that has been firmly established in ultra-peripheral collisions at RHIC, LHC, and FCC energies in the recent years.

The paper is organized as following. In Section 2, we present the production of large-$p_{T}$ charmonium and narrow resonance state (exotic charmonium) from the semi-coherent two-photon interactions in ultra-peripheral heavy ion collisions. The numerical results for proton-proton, proton-nucleus, and nucleus-nucleus collisions at RHIC, LHC, and FCC energies are plotted in Section 3. Finally, the conclusion is given in Section 4.

\section{General formalism}

The differential cross-section for the charmonium and narrow resonance state (exotic charmonium) production from the semi-coherent two-photon interaction in ultra-peripheral heavy ion collisions can be
written as
\begin{eqnarray}\label{y1}
d\sigma=\hat{\sigma}_{\gamma\gamma\rightarrow H}(W)dN_{1}dN_{2},
\end{eqnarray}
where the total cross section $\hat{\sigma}_{\gamma\gamma\rightarrow H}(W)$ for the charmonium and narrow resonance state (exotic charmonium) production from the
two-photon fusion interaction can be written in terms of the two-photon decay width of the corresponding state as\cite{25,26,27,40}
\begin{eqnarray}\label{y2}
\hat{\sigma}_{\gamma\gamma\rightarrow H}(W)=8\pi^{2}(2J+1)\frac{\Gamma_{H\rightarrow\gamma\gamma}}{M_{H}}\delta(W^{2}-M_{H}^{2}),
\end{eqnarray}
here the decay width $\Gamma_{H\rightarrow\gamma\gamma}$ for the charmonium can be taken from the experiment\cite{41}, and the narrow resonance state (exotic charmonium) can be theoretically estimated\cite{27,42,43,44}. Furthermore, $J$ and $M_{H}$ are the spin and mass of the produced state, respectively.

In the equivalent photon approximation, the effect of the electromagnetic field for an ultrarelativistic proton or nucleus is replaced by a flux of photons\cite{2,24}. Then the equivalent photon fluxes for the relativistic proton and nucleus can be obtained as
\begin{eqnarray}\label{y3}
dN_{nucleus}=\frac{\alpha Z^{2}}{\pi^{2}}\frac{1-\omega/E_{N}+\omega^{2}/2E_{N}^{2}}{\omega/E_{N}}\frac{q_{T}^{2}\left[F_{N}(q_{T}^{2}+\frac{\omega^{2}}{\gamma^{2}})\right]^{2}}{\left(q_{T}^{2}+\frac{\omega^{2}}{\gamma^{2}}\right)^{2}}d^{2}q_{T}\frac{d\omega}{\omega},\!\!\!\!\
\end{eqnarray}
\begin{eqnarray}\label{y4}
dN_{proton}=\frac{\alpha}{\pi^{2}}\frac{1-\omega/E_{p}+\omega^{2}/2E_{p}^{2}}{\omega/E_{p}}\frac{q_{T}^{2}\left[F_{p}(q_{T}^{2}+\frac{\omega^{2}}{\gamma^{2}})\right]^{2}}{\left(q_{T}^{2}+\frac{\omega^{2}}{\gamma^{2}}\right)^{2}}d^{2}q_{T}\frac{d\omega}{\omega},
\end{eqnarray}
where $\omega$ is the energy of the photon, $\gamma$ is the relativistic factor, and $F(q^{2})$ is the proton and nuclear form factor of the equivalent photon source. The energy for the projectiles are $E_{p}=\sqrt{s}/2$ for proton and $E_{N}=A\sqrt{s}/2$ for nucleus, where $A$ is the nucleon number of the nucleus, and $\sqrt{s}$ is the center of mass energy for the colliding nucleons at relativistic heavy ion collisions.

In the nucleus case, it is often used in the literature as the monopole form factor given by\cite{15,45}
\begin{eqnarray}\label{y5}
F_{N}(q^{2})=\frac{\Lambda^{2}}{\Lambda^{2}+q^{2}},
\end{eqnarray}
with $\Lambda=0.091\texttt{GeV}$ for $^{197}\texttt{Au}$ and $\Lambda=0.088\texttt{GeV}$ for $^{208}\texttt{Pb}$.

For proton projectiles, the electric dipole form factor is in general assumed to be\cite{22,22x,24}
\begin{eqnarray}\label{y6}
F_{p}(q^{2})=1/[1+q^{2}/(0.71\texttt{GeV}^{2})]^{2},
\end{eqnarray}
where $q^{2}=q_{T}^{2}+\omega^{2}/\gamma^{2}$ is the 4-momentum transfer of the proton projectile.

In the semi-coherent two-photon interaction at the nucleus-nucleus collisions, $q_{1}=(\omega_{1},\textbf{\emph{q}}_{1T},q_{1z})$ and $q_{2}=(\omega_{2},\textbf{\emph{q}}_{2T},q_{2z})\approx(\omega_{2},\textbf{\emph{0}},q_{2z})$, the total transverse momentum of charmonium or narrow resonance state (exotic charmonium) is $\textbf{\emph{p}}_{T}=\textbf{\emph{q}}_{1T}+\textbf{\emph{q}}_{2T}\approx\textbf{\emph{q}}_{1T}$ with $\textbf{\emph{q}}_{1T}\gg\textbf{\emph{q}}_{2T}$, where $\textbf{\emph{q}}_{iT}$ is the transverse momentum of the photon. Therefore the differential cross section for the nucleus-nucleus collisions can be written in the terms of charmonium and narrow resonance state (exotic charmonium) transverse momentum as the following
\begin{eqnarray}\label{y7}
d\sigma_{NN}\!\!\!\!&=&\!\!\!\!\frac{\alpha Z^{2}}{\pi^{2}}\int\frac{1-\omega_{1}/E_{N}+\omega_{1}^{2}/2E_{N}^{2}}{\omega_{1}/E_{N}}\frac{q_{1T}^{2}\left[F_{N}(q_{1T}^{2}+\frac{\omega_{1}^{2}}{\gamma^{2}})\right]^{2}}{\left(q_{1T}^{2}+\frac{\omega_{1}^{2}}{\gamma^{2}}\right)^{2}}d^{2}q_{1T}\frac{d\omega_{1}}{\omega_{1}}\nonumber\\[1mm]
&&\!\!\!\!\times\frac{\alpha Z^{2}}{\pi^{2}}\int\frac{1-\omega_{2}/E_{N}+\omega_{2}^{2}/2E_{N}^{2}}{\omega_{2}/E_{N}}\frac{q_{2T}^{2}\left[F_{N}(q_{2T}^{2}+\frac{\omega_{2}^{2}}{\gamma^{2}})\right]^{2}}{\left(q_{2T}^{2}+\frac{\omega_{2}^{2}}{\gamma^{2}}\right)^{2}}d^{2}q_{2T}\frac{d\omega_{2}}{\omega_{2}}\nonumber\\[1mm]
&&\!\!\!\!\times8\pi^{2}(2J+1)\frac{\Gamma_{H\rightarrow\gamma\gamma}}{M_{H}}\delta(W^{2}-M_{H}^{2}),
\end{eqnarray}
where the energies for the photons are $\omega_{1,2}=\frac{W}{2}\exp(\pm y)$, with $W=\sqrt{4\omega_{1}\omega_{2}}$, and the transformations $d\omega_{1}d\omega_{2}=\frac{W}{2}dWdy$ can be performed. Then we can get the differential cross-section as a function of the total transverse momentum $\textbf{\emph{p}}_{T}$ and the rapidity $y$,
\begin{eqnarray}\label{y8}
\frac{d\sigma_{NN}}{d^{2}p_{T}dy}\!=\!\frac{8\alpha^{2}Z^{4}}{\pi^{2}}(2J\!+\!1)\frac{\Gamma_{H\rightarrow\gamma\gamma}}{M_{H}}\!\!\left(\frac{2E_{N}}{M_{H}^{2}}\right)^{\!\!\!2}\!\frac{F_{N}^{2}(p_{T}^{2})}{p_{T}^{2}}\!\!\int\!\!\frac{q_{2T}^{2}F_{N}^{2}(q_{2T}^{2}\!+\!\frac{M_{H}^{4}}{16p_{T}^{2}\gamma^{2}})}{\left(q_{2T}^{2}\!+\!\frac{M_{H}^{4}}{16p_{T}^{2}\gamma^{2}}\right)^{2}}d^{2}q_{2T},\!\!
\end{eqnarray}
where $\alpha$ is the electromagnetic coupling constant, $\gamma$ is the relativistic factor, and the transverse momentum for photon is $q_{2T}>0.2\texttt{GeV}$ due to the single track acceptance condition\cite{15x,16}.

\begin{figure*}
\includegraphics[width=14cm,height=10cm]{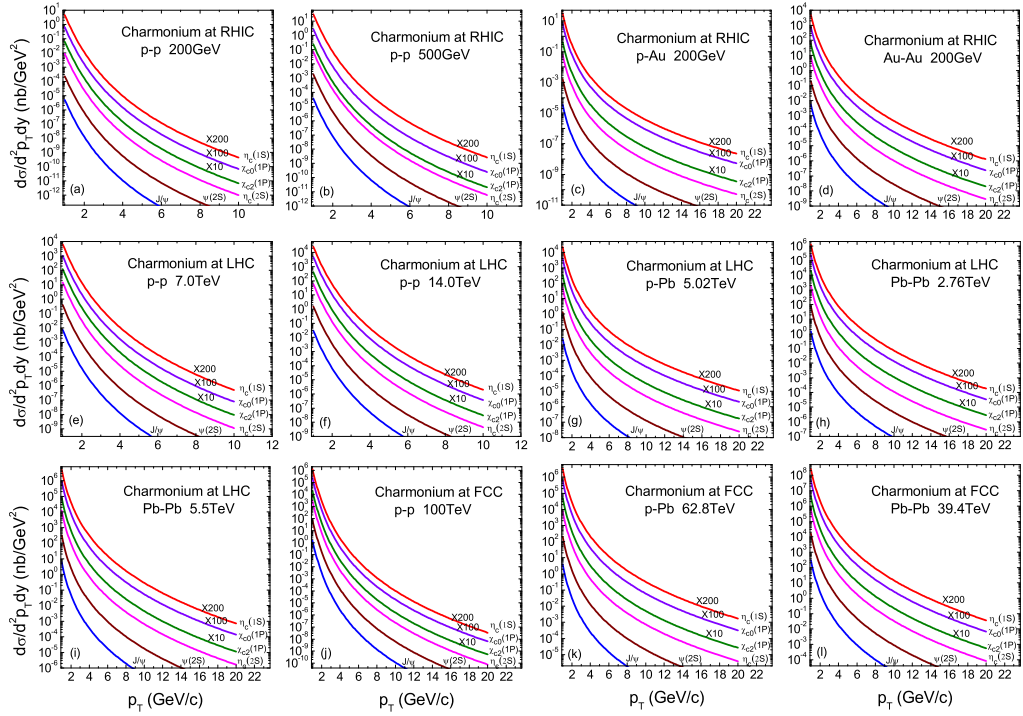}
\caption{The differential cross section for charmonium production from the semi-coherent two-photon interaction in ultra-peripheral heavy ion collisions at RHIC, LHC, and FCC.}\label{ygm1}
\end{figure*}

In the proton-nucleus collisions, the differential cross-section of the semi-coherent two-photon interactions for the charmonium and narrow resonance state (exotic charmonium) production can be written as
\begin{eqnarray}\label{y9}
\!\!\!\!\!\!\!\!\frac{d\sigma_{pN}}{d^{2}p_{T}dy}\!\!\!\!&=&\!\!\!\!\frac{8\alpha^{2}Z^{2}}{\pi^{2}}(2J\!+\!1)\frac{\Gamma_{H\rightarrow\gamma\gamma}}{M_{H}}\frac{4E_{p}E_{N}}{M_{H}^{4}}\frac{F_{N}^{2}(p_{T}^{2})}{p_{T}^{2}}\!\!\int\!\frac{q_{1T}^{2}F_{p}^{2}(q_{1T}^{2}\!+\!\frac{M_{H}^{4}}{16p_{T}^{2}\gamma^{2}})}{\left(q_{1T}^{2}\!+\!\frac{M_{H}^{4}}{16p_{T}^{2}\gamma^{2}}\right)^{2}}d^{2}q_{1T}\nonumber\\[1mm]
&&\!\!\!\!\!\!\!\!\!\!+\frac{8\alpha^{2}Z^{2}}{\pi^{2}}(2J\!+\!1)\frac{\Gamma_{H\rightarrow\gamma\gamma}}{M_{H}}\frac{4E_{p}E_{N}}{M_{H}^{4}}\frac{F_{p}^{2}(p_{T}^{2})}{p_{T}^{2}}\!\!\int\!\frac{q_{2T}^{2}F_{N}^{2}(q_{2T}^{2}\!+\!\frac{M_{H}^{4}}{16p_{T}^{2}\gamma^{2}})}{\left(q_{2T}^{2}\!+\!\frac{M_{H}^{4}}{16p_{T}^{2}\gamma^{2}}\right)^{2}}d^{2}q_{2T},
\end{eqnarray}
where the transverse momentum for the photon is $q_{iT}>0.2\texttt{GeV}$ due to the single track acceptance condition.

\begin{figure*}
\includegraphics[width=14cm,height=10cm]{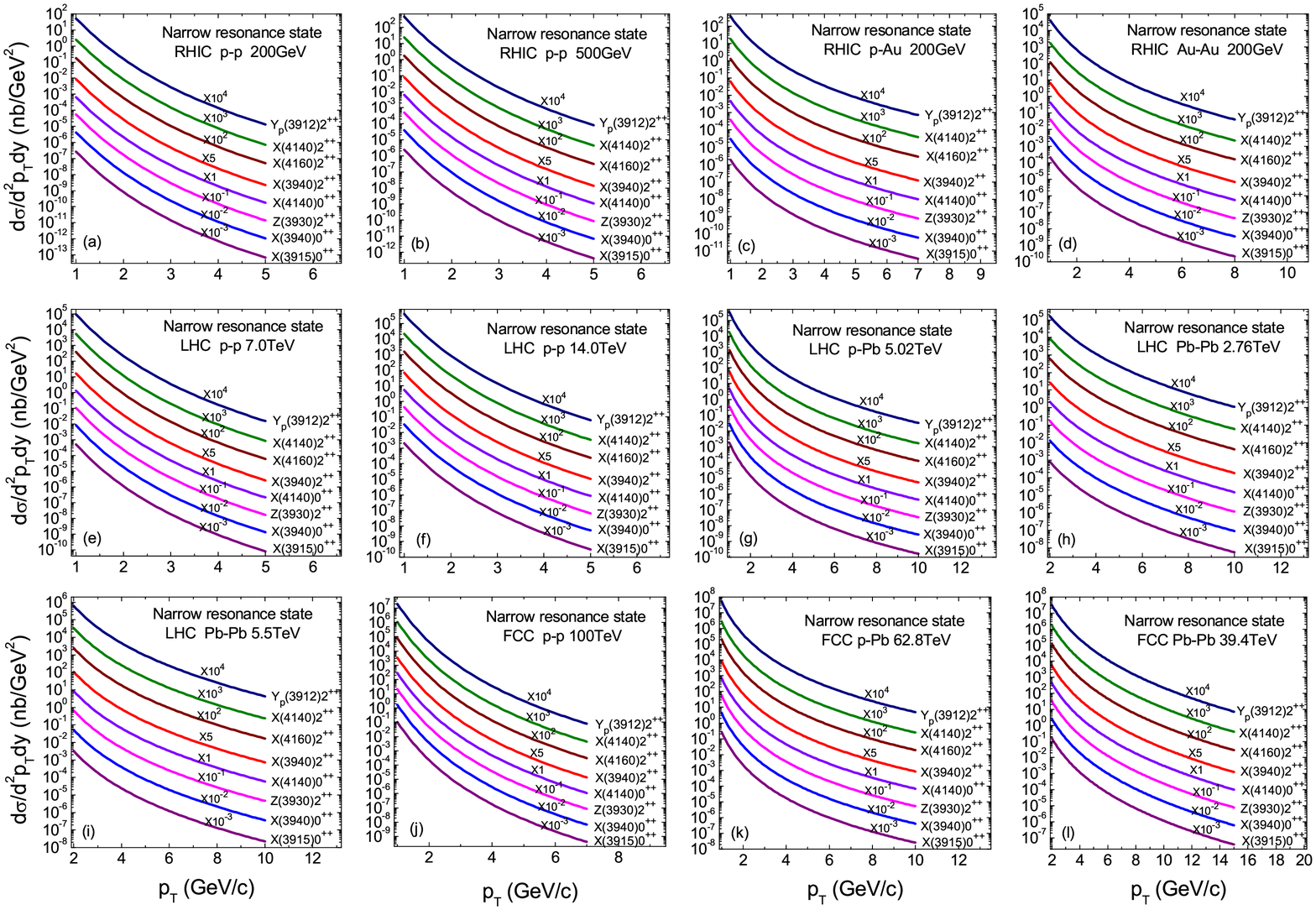}
\caption{The differential cross section for narrow resonance state (exotic charmonium) production from the semi-coherent two-photon interaction in ultra-peripheral heavy ion collisions at RHIC, LHC, and FCC.}\label{ygm2}
\end{figure*}

For the proton-proton collisions, the differential cross-section of the semi-coherent two-photon interaction for charmonium and narrow resonance state (exotic charmonium) production can be obtained as
\begin{eqnarray}\label{y10}
\frac{d\sigma_{pp}}{d^{2}p_{T}dy}\!=\!\frac{8\alpha^{2}}{\pi^{2}}(2J\!+\!1)\frac{\Gamma_{H\rightarrow\gamma\gamma}}{M_{H}}\!\!\left(\frac{2E_{p}}{M_{H}^{2}}\right)^{\!\!\!2}\!\frac{F_{p}^{2}(p_{T}^{2})}{p_{T}^{2}}\!\!\int\!\!\frac{q_{2T}^{2}F_{p}^{2}(q_{2T}^{2}\!+\!\frac{M_{H}^{4}}{16p_{T}^{2}\gamma^{2}})}{\left(q_{2T}^{2}\!+\!\frac{M_{H}^{4}}{16p_{T}^{2}\gamma^{2}}\right)^{2}}d^{2}q_{2T},
\end{eqnarray}
where the transverse momentum for photon is $q_{2T}>0.2\texttt{GeV}$ due to the single track acceptance condition.

\section{Numerical results}

The equivalent photon fluxes for the proton and nucleus become very large at the RHIC, LHC, and FCC energies. The photon flux is high enough that two-photon interactions may be accompanied in relativistic heavy ion collisions. Especially for the nucleus-nucleus collisions, the photon flux scales as $Z^{2}$, and the two-photon differential cross-section scales as $Z^{4}$. Moreover, in a semi-coherent approach, we keep the transverse momentum of one of the photons small, then the whole nucleus acts coherently. In Figs. 1 and 2, we plot the differential cross section for large-$p_{T}$ charmonium [$J/\psi$, $\psi(2s)$, $\eta_{c}(1S)$, $\eta_{c}(2S)$, $\chi_{c0}(1P)$, and $\chi_{c2}(1P)$] and narrow resonance state [$X(3940)$, $X(4140)$, $Z(3930)$, $X(4160$), $Y_{p}(3930)$, and $X(3915)$] production in proton-proton, proton-nucleus, and nucleus-nucleus collisions in ultra-peripheral heavy ion collisions. For proton-proton and proton-nucleus collisions, our results for the production of charmonium and narrow resonance state (exotic charmonium) are not prominent at RHIC and LHC energies. But the predicted differential cross sections increase with the energy, and it can not be negligible at FCC energies. For nucleus-nucleus collisions, the differential cross sections are enhanced by a factor of $Z^{4}$ ($Z^{2}$) in comparison to proton-proton (proton-nucleus) collisions in our calculations, and the contribution of the semi-coherent two-photon interactions for large-$p_{T}$ charmonium and narrow resonance state becomes evident in the ultra-peripheral heavy ion collisions at RHIC, LHC, and FCC energies. Indeed, the main sources of changes in the differential cross sections are the magnitude of the decay width and the spin of the produced particle, since the masses of the narrow resonance states are nearly the same. Here we observe a smaller difference of the differential cross sections between the narrow resonance states (exotic charmonium).

\section{Conclusion}

In summary, we have investigated the production of large-$p_{T}$ charmonium and narrow resonance state (exotic charmonium) from the semi-coherent two-photon interactions in proton-proton, proton-nucleus, and nucleus-nucleus collisions in ultra-peripheral heavy ion collisions. In the equivalent photon approximation, the effect of the electromagnetic field for the ultrarelativistic proton and nucleus is replaced by the flux of photons. Then, we show the transverse momentum distribution of the charmonium and narrow resonance state (exotic charmonium) production by using the monopole form factor of nucleus and the electric dipole form factor of proton. Our calculations show that the large values of the differential cross sections for nucleus-nucleus and proton-nucleus collisions, and non-negligible values in proton-proton collisions can be obtained with the semi-coherent approach in ultra-peripheral heavy ion collisions at the RHIC, LHC, and FCC energies.

\section{Acknowledgements}

This work is supported by the National Basic Research Program of China (973 Program)
with Grant No. 2014CB845405 and the National Natural Science Foundation of China with
Grant No. 11465021 and No. 11065010.

\end{document}